\documentclass[twocolumn,aps]{revtex4-1}

\usepackage{graphicx}
\usepackage{amssymb}
\usepackage{amsfonts}
\usepackage{amsmath}
\usepackage{amsbsy}
\usepackage{natbib}
\usepackage{comment}
\usepackage{color}

\renewcommand{\vec}[1]{\mbox{\boldmath$\mathrm{#1}$}}
\newcommand{\be}{\begin{equation}}
\newcommand{\ee}{\end{equation}}
\newcommand{\ben}{\begin{eqnarray}}
\newcommand{\een}{\end{eqnarray}}

\begin{document}

\title{Thermally induced magnonic spin current, thermomagnonic torques and domain wall dynamics in the presence  of Dzyaloshinskii-Moriya interaction }

\author{X.-G. Wang$^{1,2}$}

\author{L. Chotorlishvili$^{1}$}
\email{levan.chotorlishvili@gmail.com}

\author{G.-H. Guo$^2$}

\author{A. Sukhov$^{1}$}

\author{V. Dugaev$^{3}$}

\author{J. Barna\'{s}$^{4}$}

\author{J. Berakdar$^{1}$}

\affiliation{
$^{1}$Institut f\"ur Physik, Martin-Luther Universit\"at Halle-Wittenberg, 06099 Halle (Saale), Germany\\
$^{2}$School of Physics and Electronics, Central South University, Changsha 410083, China \\
$^{3}$Department of Physics and Medical Engineering, Rzeszow University of Technology, 35-959 Rzeszow, Poland\\
$^{4}$Faculty of Physics, Adam Mickiewicz University, 61-614 Poznan, Poland
}

\begin{abstract}
Thermally activated domain wall (DW) motion in magnetic insulators has been considered theoretically, with a particular focus on the role of Dzyaloshinskii-Moriya Interaction (DMI) and thermomagnonic torques. The thermally assisted DW motion is a consequence of the magnonic spin current due to the applied thermal bias. In addition to the exchange magnonic spin current and the exchange adiabatic and the entropic spin transfer torques, we also consider the DMI-induced magnonic spin current, thermomagnonic DMI field-like torque and the DMI entropic torque. Analytical estimations are supported by  numerical calculations. We  found that the DMI has a substantial influence on the size and the geometry of DWs, and that the DWs become oriented parallel to the long axis of the nanostrip. Increasing the temperature smoothes   the DWs. Moreover, the thermally-induced magnonic current generates a torque on the DWs, which is responsible for their motion. From our analysis it follows that for a large enough DMI the influence of DMI-induced field-like torque is much stronger than that of the DMI and the exchange entropic torques. By manipulating the strength of the DMI constant, one can control the speed of the DW motion, and the direction of the DW motion can be switched, as well. We  also found that DMI not only contributes to the total magnonic current, but also it modifies the exchange magnonic spin current, and this  modification depends on the orientation of the steady state magnetization. The observed phenomenon  can be utilized in spin caloritronics devices, for example in the DMI based thermal diodes. By switching the magnetization direction, one can rectify the total magnonic spin current.
\end{abstract}

\date{\today}

\maketitle

\section{Introduction}

Magnetic domains in ferromagnetic materials are separated by domain walls (DWs). One may attribute DWs to the spontaneously broken symmetry in the system. Formally a DW pretty much resembles a kink solution of the Sine-Gordon model. The existence of domains and DWs is promoted energetically. Finite width of the DWs softens the transition between differently magnetized regions
and minimizes the total system's energy. DWs are currently attracting growing interest from both fundamental and application points of view \cite{AlXi05,HaTh08,PaHa08,PaYa15}. A particular proposal is to exploit DWs for high density storage in a "racetrack" shift memory \cite{PaHa08,PaYa15}.

It is well known, that the torque exerted by a spin-polarized current can drag DWs \cite{YaNa04,ZhLi04,ThNa05}. The velocity of the DW's motion is an important parameter when it comes to applications but
energy consumption and economically justifiable operating costs are also pertinent issues in this respect.
In magnetic insulators, the free charge  carriers are absent. As information carrier,  the magnonic spin current induced by an applied temperature gradient and the spin Seebeck effect \cite{UcXi10} could serve as an alternative  \cite{RiHi15}. Due to the low energy consumption, thermally induced DW dynamics has certain advantage \cite{BaSa12}.

The thermally induced DW dynamics has been studied intensively in recent years \cite{ToMa12,JiUp13,ChBe14,ScRi14,WaWa14}.
The effect of applied thermal bias is twofold: (i) it generates the adiabatic thermomagnonic spin transfer torque driving DWs from the hot to the cold edge, \cite{HiNo11,EtCh14,KiTs15}, and (ii) it forms the non-adiabatic thermomagnonic torque, also known as entropic torque or dissipative torque.  The non-adiabatic thermomagnonic torque is quantified in terms of the local magnon density and is related to the temperature dependence of the effective exchange field \cite{KiTs15}. The entropic torque tends to move DWs towards the hot edge. This effect is predetermined thermodynamically due to the DW's free energy potential landscape. The above scenario is appropriate when magnons are completely transmitted through the wall. The situation is different when magnons are  reflected from DWs.  Microscopic theory anticipates   a strong magnonic recoil effect for the DW motion \cite{YaCa15}.
If magnon is reflected by a  DW, then linear momentum is transferred to the DW\cite{WaGu12}, and the  DW is pushed towards the cold edge.

Recently a nontrivial role of the Dzyaloshinskii-Moriya interaction (DMI) in the context of thermally activated magnonic currents was invoked. The DMI originates from  structural inversion asymmetry and spin-orbit coupling. Thermally driven magnonic current in the presence of  DMI generates field-like  and damping-like torques \cite{MaNd14}. The role of the DMI in the formation of the reactive and dissipative thermomagnonic torques has been studied in \cite{KoGu15}.  DMI can also change the direction and the speed of the magnonic current and it generates a magnonic momentum transfer torque \cite{WaAl15}. This torque can drastically influence the motion of the DW in a nanowire.

The diversity of the effects caused by DMI naturally enriches the physics and possible scenarios of the DW's motion. However, these effects are not precisely categorized in the sense of their cumulative impact on the DW's motion, despite  the huge interest in DWs, \cite{KoGu15,ThRo12,BoRo13,MaEm14,WaAl15,TrAb10}. Therefore, in the present work we intend to build up a comprehensible
description of this impact.  Our main interest concerns the role of DMI in the thermally assisted motion of DWs, and our conclusions are based on  analytical estimations and also on micromagnetic simulations. For completeness
we consider two diverse geometries of the samples: 3D nanostrip  and 1D nanowire. We show that thermomagnonic torques related to DMI are apparently stronger than the exchange thermomagnonic torques. Therefore, the DMI thermomagnonic torques dominate in the thermally assisted DW motion. We start our considerations from micromagnetic simulations of domain structure in a nanostrip. Then, we consider analytically and numerically the 1D nanowire. In section 2 we present results of numerical simulations of magnetic  domains and DWs. Spin currents due to exchange and DMI in a magnetic nanowire are presented in section 3.  Exchange and  DMI thermomagnonic torques are calculated in section 4, while DW motion is described in section 5. Summary and final conclusions are  in section 6.

\section{Effect of DMI on magnetic domains and DWs}

In this section we consider the impact of DMI on the domain structure and the domain walls in a 3D nanostrip, and show numerically that  DMI  substantially influences the formation of the domains.   Our calculations show that DMI modifies the width of the DWs as well as the size, the shape and the arrangement of the domains. This is because DMI favors nonuniform magnetization.

We consider bulk DMI with the corresponding energy density $\zeta_{DMI}=D\vec{m}\cdot\big(\nabla\times\vec{m}\big)$, where $D$ is the parameter of DMI and $\vec{m}$ is a unit vector along the magnetization, $\vec{m}=\vec{M}/|\vec{M}|$. The uniaxial anisotropy is assumed along the $x$ and $z$ axis and is defined as  $-K_xm_{x}^{2}$ and $K_zm_{z}^{2}$, respectively, where $K_x$ and $K_z$ are the relevant anisotropy constants. The contribution of the exchange interaction, DMI, and magnetic anisotropy to the total free energy of the system reads:
\begin{equation}
\displaystyle E= \int\big[A\big(\nabla\vec{m}\big)^{2}-K_{x}m_{x}^{2}+K_{z}m_{z}^{2}+\zeta_{DMI}\big] d\vec{r},
\label{eq_Hamiltonian}
\end{equation}
where $A$ is the exchange stiffness parameter. Note, the code used for numerical simulations also includes the dipolar magnetostatic energy.

In the numerical calculations described below we assumed the following material parameters: $M_S = 3.84 \cdot 10^5$ A/m for the saturation magnetization $M_S$, $A = 8.78 \cdot 10^{-12}$ J/m, $K_x = 1 \cdot 10^5$ J/m$^3$, $K_z =- 2 \cdot 10^5$ J/m$^3$ and $\alpha= 0.05$ for the Gilbert damping parameter.
Strength of the DMI  is varied within the interval  $-1.58$ mJ/m$^2$ $\leq D\leq$  $1.58$ mJ/m$^2$. The material parameters we used are typical of FeGe, but the results obtained are of a  general nature and not limited to a certain material only. For example, because of the sizeable magnetoelectric coupling in the one phase chiral multiferroic systems, \cite{WaNe03,YaMi06,MeMo12,AzCh14} strength of the DMI can be controlled by means of the applied external electric field.

\begin{figure}[htb]
\centering
\includegraphics[width=0.42\textwidth]{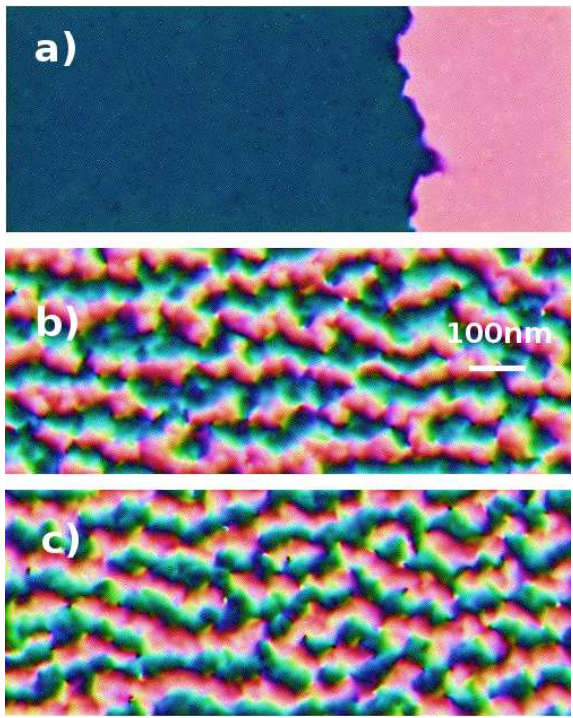}
\caption{Magnetization configuration of a $1000$ nm $\times 400$ nm $\times 10$ nm nanostrip for a) $D=0$ mJ/m$^2$ after a relaxation of $310$~ns; b) $D=-1.58$ mJ/m$^2$ and c) $D=1.58$ mJ/m$^2$  after the relaxation towards equilibrium ($510$~ns). In all cases the temperature $T=0$ is constant in the system (zero temperature gradient). The color coding indicates the  out-of-plane magnetization orientation -- towards the reader (red) and away from the reader (dark blue). Simulations have been performed using the \texttt{mumax3}-simulation package \cite{mumax3}.}
\label{fig_1c}
\end{figure}

In numerical simulations, the initial magnetization configuration of a $1000$ nm $\times 400$ nm $\times 10$ nm nanostrip was random and then relaxed to the ground state magnetic configuration.   In Fig.1 we show the results of the simulations  for zero temperature (no temperature gradient) in the whole system, $T=0$.  In the absence of DMI, two relatively large domains are formed in the nanostrip, see Fig.~\ref{fig_1c}a. DMI leads to a fragmentation of the large
 domains into the horizontal pattern of thin domains,
as shown in Fig.~\ref{fig_1c}b and Fig.~\ref{fig_1c}c. This fragmentation appears since the DMI favors nonuniform magnetization.
The individual domains are separated by DWs, whose width and geometry plays a decisive role in their motion. The mass of the thin DWs is smaller and therefore their mobility is higher. Apart from this,
due to the changes in the geometry of the domains, the surface of the horizontal thin DWs is larger. This  can increase effectively the interaction of the DWs with the magnonic spin current (cf.  Fig.~\ref{fig_1c} corresponding to the case when the  magnonic current is absent).

\begin{figure}[htb]
\centering
\includegraphics[width=0.42\textwidth]{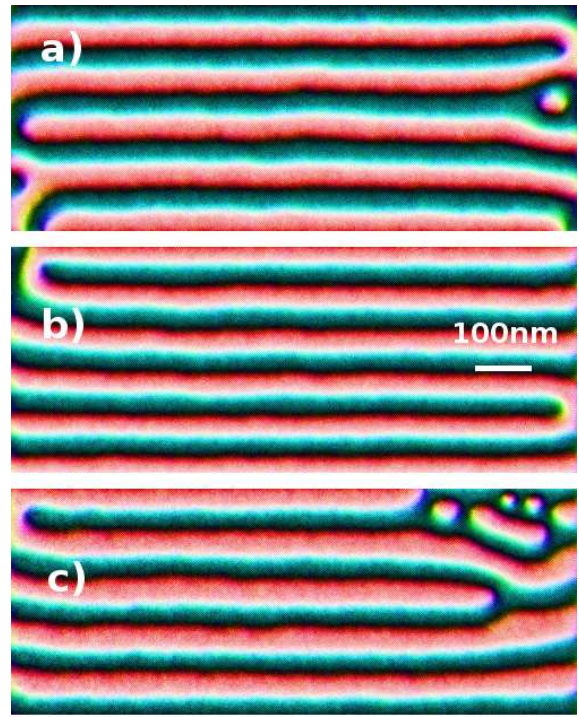}
\caption{Equilibrium configuration of a $1000$ nm $\times 400$ nm $\times 10$ nm nanostrip for a) $D=-1.58$ mJ/m$^2$ and  b) $D=1.58$ mJ/m$^2$, both at $T=300$~K and in the absence of temperature gradient; and c) $D=1.58$ mJ/m$^2$ in the presence of a linear  temperature gradient, $d T=0.3$~K/nm. The color coding stands for the  out-of-plane magnetization direction -- towards the reader (red) and away from the reader (dark blue). The average domain wall thickness is around $42$~nm. Simulations have been performed using the \texttt{mumax3}-simulation package \cite{mumax3}.}
\label{fig_1d}
\end{figure}

The DWs become smoother at higher temperatures, and the domains become parallel to the long axis of the strip. This is shown in Fig.2a and Fig.2b, where $T=300$K while the other parameters are the same as in Fig.1b and Fig.1c, respectively. In Fig.2c there is additionally a linear temperature gradient in the system, $T(x)=-d T\,x+450$K, while the other parameters are as in Fig.2b. We assumed that the middle of the strip is at $x=0$, so the left and right ends of the strip correspond to $x=-500$nm and $x=500$nm, respectively.  For the assumed $d T=0.3$K/nm, the temperatures of the left and right ends of the strip are 600K and 300K, respectively. This externally applied  thermal bias generates a magnonic spin
current. A flux of magnons is oriented from the hot to the cold edge of the sample (from left to right). However, the direction of the associated  spin current may be different. If the equilibrium magnetization component is positive (in our case $m_{x}>0$, see below), then the magnon's spin projection is positive (equal to $\hbar$), and the direction of the spin current is parallel to that of the magnon flux. If  $m_{x}<0$, then the magnon's spin projection is negative and the spin current flows in the opposite direction to the magnon current.
\begin{figure}[htb]
\centering
\includegraphics[width=0.42\textwidth]{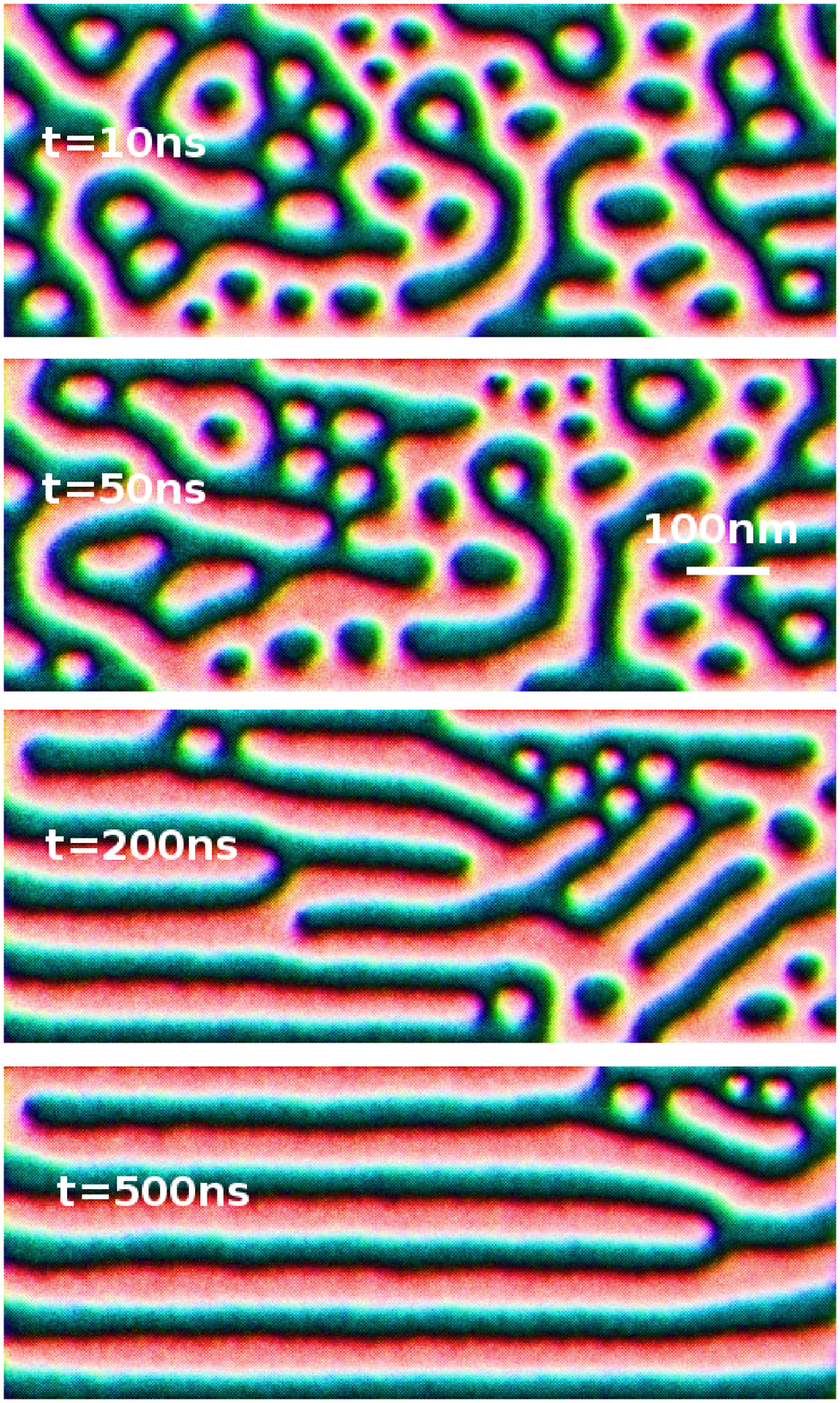}
\caption{Time-resolved domain wall propagation for a $1000$ nm $\times 400$ nm $\times 10$ nm nanostrip under the thermal bias corresponding to  $d T=0.3$~K/nm and for $D=1.58$ mJ/m$^2$. Red/dark blue: out-of-plane magnetization orientation pointing to/away from   the reader.}
\label{fig_1f}
\end{figure}

On the  way to the cold edge, the magnonic current encounters the DWs. Our previous studies \cite{Arthur} show that depending on the DW's width, the  magnonic current is either transmitted
through a DW (thin DWs) or it is reflected by the DW (thick DWs). Reflected magnonic current exerts a certain pressure on the DW's surface and pushes the DW to the cold edge. The pressure exerted by the magnonic current on a DW is: $P=2\delta nv_{k}\hbar k$. Here $v_{k}=\mid\vec{v_{k}}\mid$ and $k=\mid \vec{k}\mid$ are the velocity and the wave vector of magnons, respectively, while $\delta n=n_{\rm neq}\big(T\big)-n_{\rm eq}\big(T\big)$ quantifies the magnon accumulation effect, i.e. the excess of the density of non-equilibrium magnons $n_{\rm neq}\big(T\big)$ compared to the reference number of equilibrium magnons $n_{\rm eq}\big(T\big)$ at the same temperature, but in the absence of thermal gradient. In turn, if the magnon current passes through the DW, then the dominant factor in the DW's motion is the entropic torque. The free energy  $F = \bar E - TS$ (here $\bar E$ is the internal energy and $S$ is the entropy) is minimized at elevated temperatures $T$. Thus, the entropic torque drives
DWs towards the hotter edge.

The magnonic current can also modify significantly the shape of the domains and DWs. As we see from Fig.~\ref{fig_1d}c, the magnonic current  moves all inhomogeneities to the colder end of the strip, which appears as a result of the interaction between the magnonic  current and DWs due to the exchange coupling and DMI.
The time evolution of the domain reconstruction and the DW motion due to a thermally-induced magnon current
is shown in Fig.~\ref{fig_1f}. The initial magnetic configuration (at $t=0$) was assumed to be random, and then a thermal bias was applied. Because of the spin current that is driven by the temperature gradient, the domains start to move. Fig.3a shows the domain pattern after the first 10ns. One can see, that after 500 ns (cf. Fig.3d) the domain pattern reaches the stationary state also shown in Fig.~\ref{fig_1d}c.

From the above follows, that one of the interesting effects of the DMI is the rearrangement and the reconstruction of the magnetic domains and DWs. In the absence of DMI,  the DWs are typically orthogonal to the nanowire axis (and magnonic spin current). 
As a result of DMI, DWs become oriented mostly parallel to the long axis of the strip (and the applied thermal bias). Thus, instead of exerting a pressure on DWs, the magnonic current drags them towards the cold edge due to  the 'viscosity' resulting from the exchange interaction and DMI. Reconstruction of the DWs to align  parallel to the $x$ axis minimizes the total energy of the system.
 However, in the general 3D case the structure of the system is quite complex and no analytical estimation can be achieved. Therefore, for the sake of simplicity in what follows we will consider a  1D nanowire.

\section{Exchange and DMI spin currents}

We consider now a 1D nanowire along the $x$ axis. A uniform equilibrium local magnetization $\vec{M}$ is assumed to be oriented along the $x$ axis, which can be either positive, ${M}_{x}=M_S>0$, or negative, ${M}_{x}=-M_S<0$. This holds in the low-temperature limit and when the magnetic anisotropy and the exchange interaction dominate over the DMI.
Higher temperatures activate the magnetization dynamics leading to slight changes in the longitudinal magnetization $M_{x}$ and to nonzero transversal $M_{y},M_{z}$ components. In the presence of temperature gradient, the longitudinal component of the magnetization, $M_{x}$, is generally not uniform and it depends on the local magnon temperature in the nanowire. The nonuniform temperature profile in the sample is quantified {\it via} the strength of the noise implemented into the stochastic Landau-Lifshitz-Gilbert (LLG) equation. In this section, however,  the longitudinal component of the magnetization is irrelevant (except its sign), and therefore we can assume it as a constant. The main role is played by the transversal components $M_{y},M_{z}$, which undergo random thermal agitation. These magnetic excitations can be described by
$\vec{M}_{sw}=(0,M_y(x,t),M_z(x,t))$.

The spin current in magnetic insulators is solely mediated by magnons. In the following we decompose this spin current into two terms: exchange and DMI ones.
In fact, the formula and the numerical results given below correspond not to spin currents but to magnetization currents.
These are not equivalent though related. Each magnon carries a spin momentum of magnitude $\hbar$ and opposite magnetic moment of magnitude $m_B$, with $m_B$ standing for the Bohr magneton.
For simplicity, in the following we will refer to them interchangeably as spin currents or magnetization currents, bearing in mind that the spin current has opposite sign to that of the corresponding magnetization current.

Contribution due to the exchange interaction to the magnetization current density $\vec{J}_A$ {\it via} magnons can be calculated from the formula~\cite{KaHa10}
\begin{equation}
\displaystyle \partial_x \vec{J}_A = \big \langle \gamma \vec{M}_{sw} \times \vec{h}_{A} \big \rangle .
\label{eq_1}
\end{equation}
Here, $\vec{h}_{A}=2A\partial^2_x {\bf M}_{sw}/(\mu_0 M^2_S)$ is the exchange field, $\gamma =\mu_0 e/2m$ is the gyromagnetic ratio, and $\mu_0$ is the magnetic permeability constant. After substituting $\vec{M}_{sw}$ into Eq. (\ref{eq_1}), the $x$ component of the exchange magnonic spin current, $J_A^x$, can be written as
\begin{equation}
\displaystyle J_A^x = l_A \big \langle M_y \partial_x M_z - M_z \partial_x M_y\big >,
\label{eq_2}
\end{equation}
where $l_A = 2 \gamma A / (\mu_0 M^2_S)$. Other components of the exchange spin current are zero due to the geometry of the 1D nanowire and the magnetization ordering. Note, the magnetization current density is defined as the magnetic moment flowing through a unit area per unit time, and is measured in the units of A/s.

Similarly, the DMI-induced magnetization current density, $\vec{J}_{\rm DMI}$, can be obtained from the formula
\begin{equation}
\displaystyle \partial_x \vec{J}_{\rm DMI} = \big \langle \gamma \vec{M}_{sw} \times \vec{h}_{\rm DMI} \big \rangle ,
\label{eq_3}
\end{equation}
where $\vec{h}_{\rm DMI}=-2D (\vec{\nabla} \times \vec{M}_{sw})/(\mu_0 M^2_S)$ is the dynamical DMI field and $D$ is the bulk type DMI constant. From Eq. (\ref{eq_3}) we deduce the following formula for the  $x$-component of the DMI magnetization current density, $J^x_{\rm DMI}$:
\begin{equation}
\displaystyle J^x_{\rm DMI} = -\frac{1}{2}l_D \big \langle M_y^2 + M_z^2 \big \rangle ,
\label{eq_4}
\end{equation}
where $l_D = 2 \gamma D / (\mu_0 M^2_S)$. The other components of $\vec{J}_{\rm DMI}$ are zero for the same reasons as in the case of the exchange spin current.
We note, that in the ground state  magnetic configuration we have  ${M}_{x}=M_S$, and thus ${M}_{y}={M}_{z}=0$.
Therefore, the ground state average of the magnonic spin currents is zero:  $\big \langle M_y \partial_x M_z - M_z \partial_x M_y\big > =0$,  $\big \langle M_y^2 + M_z^2 \big \rangle = M_{s}^{2}-\big \langle M_x^2 \big \rangle=0$. This situation changes when a thermal bias is applied.
 At a finite temperature, the mean value of the magnetic moment parallel to the $x$ axis can be estimated as
  ${M}_{x}=L\bigg(\frac{M_{x}H_{\rm eff}}{k_{B}T_{\rm mg}}\bigg)$. Here $L(...)$ is the Langevin function, $H_{\rm eff}$ is the effective magnetic field, $k_{B}$ is the Boltzmann constant, and $T_{\rm mg}$ is the effective magnon temperature.
   A nonzero net spin current in the system may exist only if ${M}_{x}<M_S$.
   This inequality holds when the magnon temperature $T_{\rm mg}$ is finite.

Using the above definitions one can calculate the  magnonic spin current (magnetization current) density in magnetic insulators, provided the dynamical components $M_y$ and $M_z$ are known. Upon implementing the linear thermal gradient, these components can be calculated from the stochastic LLG equation
\begin{equation}
\displaystyle \partial_t \vec{M} = - \gamma \vec{M} \times \left( \vec{H}_{\rm eff} + \vec{h}_l\right) + \frac{\alpha}{M_S}\vec{M} \times \partial_t \vec{M},
\label{eq_5}
\end{equation}
where  $\alpha$ is the phenomenological Gilbert damping constant, and the expression for the  effective magnetic field $\vec{H}_{\rm eff}$ reads:
\begin{equation}
\displaystyle \vec{H}_{\rm eff} = \frac{2}{\mu_0 M^2_S}\left(A\nabla^2\vec{M}-D\vec{\nabla} \times \vec{M}+ K_x M_x \vec{e}_x-K_z M_z \vec{e}_z\right),
\label{eq_6}
\end{equation}
where the first term corresponds to the exchange field,
the second term describes the DMI contribution, while the last two terms stand for the uniaxial $K_x$ and the easy xy-plane $K_z$ anisotropy fields. The random magnetic field $\vec{h}_l$ is characterized by the correlation function of the white noise
\begin{equation}
\displaystyle \big \langle h_{l,i}(\vec{r},t) h_{l,j}(\vec{r'},t)\big \rangle = \frac{2k_B T(\vec{r}) \alpha}{\gamma M_S}\delta_{ij}\delta(\vec{r}-\vec{r'})\delta(t-t'),
\label{eq_7}
\end{equation}
where $k_B$ is the Boltzmann constant and $T(\vec{r})$ is the 
local temperature.

\begin{figure}[htb]
\centering
\includegraphics[width=0.49\textwidth]{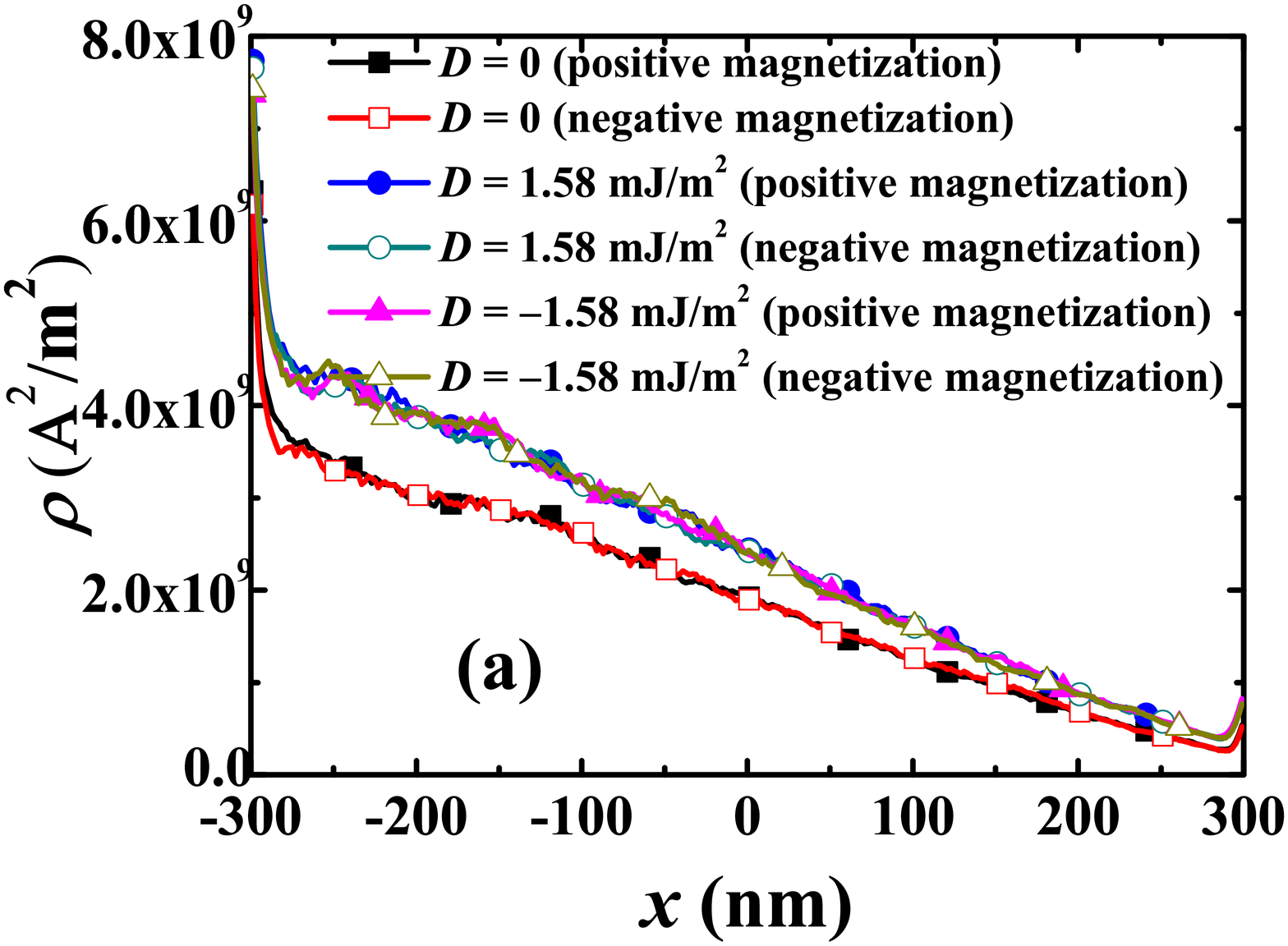}
\includegraphics[width=0.49\textwidth]{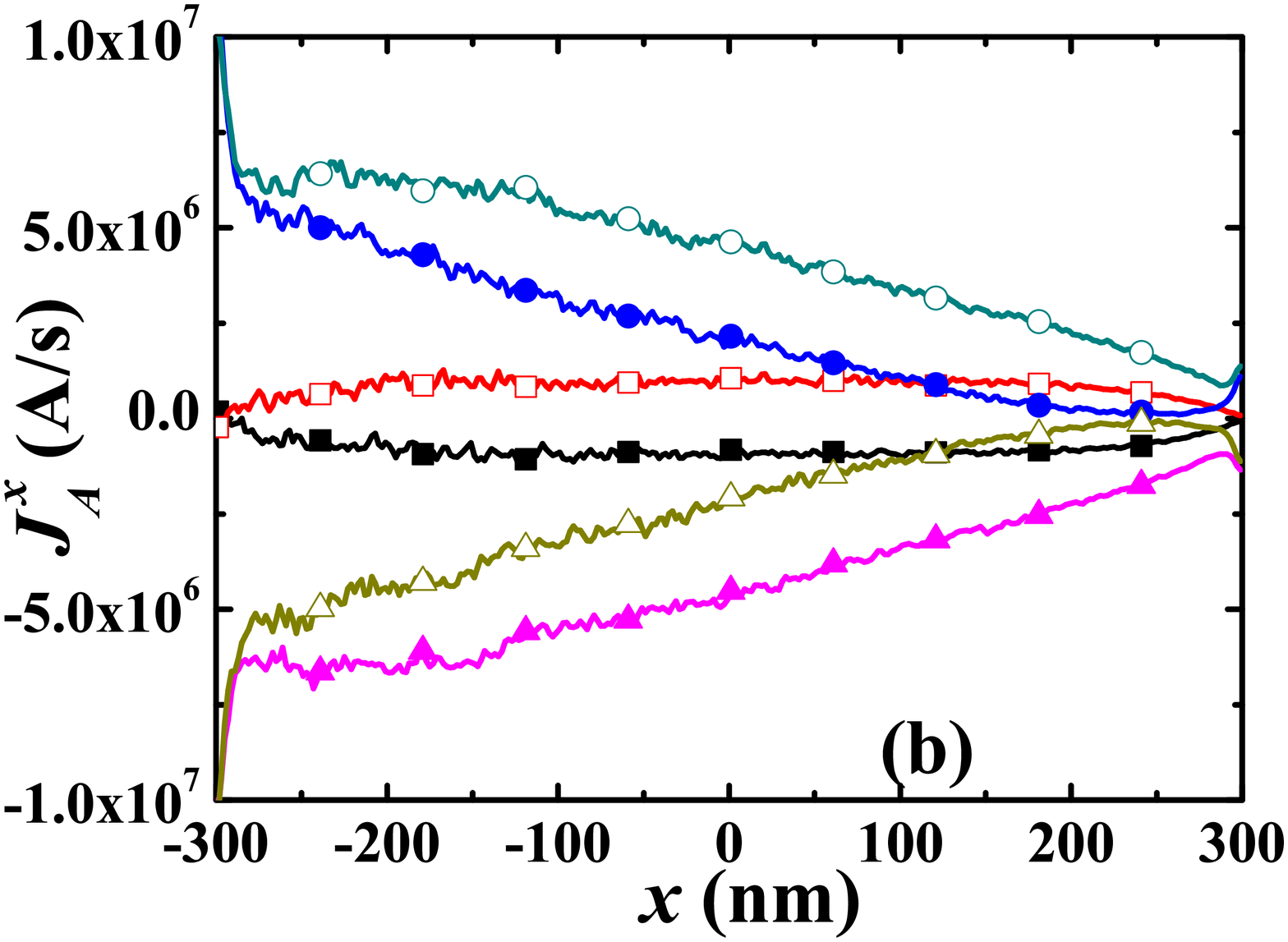}
\caption{Profile of the local magnon density $\rho$ (a) and the exchange magnonic spin current $J^A_x$ (b) in the nanowire aligned along the x axis for the parameters $D = 0$ (squares), $D = +1.58$ mJ/m$^2$ (circles), $D = -1.58$ mJ/m$^2$ (triangles), and for  positive (solid dots) and a negative (open dots) magnetization. The applied thermal bias was $dT=0.03$K/nm. }
\label{fig_1}
\end{figure}

For the numerical calculations we assumed the length of the nanowire  to be 600 nm and the unit cell size of 2 nm. A linear temperature gradient,
$T(x) = -dT (x - 300 nm)$ has the slope $dT$.   Thus, for $dT=0.03$K/nm we have T=18K at the left side, $x_L=-300$nm, and T=0K at the right side, $x_R=300$nm. The material parameters used in the numerical calculations were the same as those used in the preceding section.
We have considered both positively, $m_{x}>0$, and negatively, $m_{x}<0$,  magnetized nanowires. In the absence of DMI we observed (see Fig. \ref{fig_1}a) a linear gradient of the local magnon density. This result is in agreement with the previous studies \cite{KiTs15,HoSa13,Kova14}. The local magnon density is quantified here via the squared transversal magnetization components averaged over time, $\rho = \langle M_y^2 + M_z^2 \rangle $. Note, $\rho$  is then measured in the units of (A/m)$^2$.   In both cases (i.e. for positive and negative magnetization), $\rho$ decreases linearly with $x$ and thus also with the temperature $T$, as shown in Fig.~\ref{fig_1}(a). Some deviations from the linear dependence
at the boundaries appear due to magnon reflections from the edges.  We have also calculated the local magnon density $\rho$ for different strengths of the DMI constant ($D = \pm 1.58 \cdot 10^{-3}$ J/m$^2$) and different directions of the magnetization,  $m_{x}>0$ and $m_{x}<0$. We observed  no qualitative changes in the magnon density. For $m_{x}>0$ and $m_{x}<0$ the dependence on $x$ is linear (see Fig. \ref{fig_1}(a)). However, the DMI  increases remarkably the magnon density $\rho$ as compared to the case of $D = 0$.

In Fig.~\ref{fig_1}(b) we plot the exchange-induced magnetization current density, $J_A^x$. Consider first the limit of $D=0$, i.e., the absence of DMI.  When the magnetization is positive, $m_{x}>0$,
then  $J_A^x$ is negative (note the true spin current is then positive). Exchange  magnons flow along the axis $x$ from the hot edge to the cold one. For a negative magnetization, the exchange magnons still flow from the hot edge to the cold one, but the magnetization current $J_A^x$ is then positive, as shown in Fig. \ref{fig_1}(b). This is obvious because the spin of a magnon is then negative ($-\hbar$), while the magnetic moment is positive. We note that the current $J_A^x$ is not uniformly distributed along the wire, and the maximum of the absolute value of $J_A^x$  is found in the middle of the nanowire, while the spin current densities at the boundaries are rather small. This effect has clear explanation in terms of the local exchange spin torque~\cite{EtCh14}.

Expression for  the magnetization current tensor, Eq.(\ref{eq_1}), can be presented in the recursive discrete form:
\begin{equation}
\label{longitudinalspincurrent}
 \displaystyle I^{\alpha}_{n}=I^{\alpha}_{0}-\frac{2Aa}{M_{S}^{2}}\sum_{m=1}^{n}m^{\mathrm{\beta}}_{\mathrm{m}}(m^{\mathrm{\gamma}}_{\mathrm{m-1}}+m^{\mathrm{\gamma}}_{\mathrm{m+1}})\varepsilon_{\alpha\beta\gamma},
\end{equation}
where $\varepsilon_{\alpha\beta\gamma}$ is the Levi-Civita antisymmetric tensor, Greek indexes define the current components and the Latin ones denote sites  (elementary cells) of the nanowire, $n,m=1,...N$ with $N$ being the total number of elementary units (assumed  even for simplicity).
The local exchange spin torque is defined through the following relation~\cite{EtCh14}:  $Q_{n}=-\big \langle \gamma \vec{m}_{n} \times \vec{h}_{A,n} \big \rangle$, where $\vec{h}_{A,n}$ is the local exchange field acting on the magnetic moment of the $n$-th  cell,  and the spin current tensor is related to the spin current tensor density via $I_{n}^{s}=a^{2}J_{n}^{s}$, ($a$ is the size of unit cell). Then, taking into account the local exchange spin torque and the recursive Eq.(\ref{longitudinalspincurrent}) we deduce: $I_{n}=I_{n-1}+\frac{a^3}{\gamma}Q_{n}$. As we see, the  spin current is increasing from one site to the next site when the local exchange torque is positive, $Q_{n}>0$. If a linear thermal bias is applied, $T_{L}>T_{R}$, the local magnonic temperature in the center of the nanowire is equal to $T_{N/2}^{m}=\big(T_{R}+T_{L}\big)/2$ (the magnon temperature profile formed in the nanowire is linear, too). On the left to the center, the  magnon temperature is higher than $T_{N/2}^{m}$, and extra magnons are produced. The local exchange torque is positive, $Q_{n}>0$, and the magnonic spin current is increasing until the center of the nanowire is reached. To the right of the center the situation is different; the temperature is lower than the average temperature. Instead of producing magnons those  are absorbed (for the details of the magnon accumulation effect we refer to \cite{Hinzke3}), the  local exchange torque becomes negative  $Q_{n}<0$ and the magnonic spin current decreases.

\begin{figure}[htb]
\centering
\includegraphics[width=0.49\textwidth]{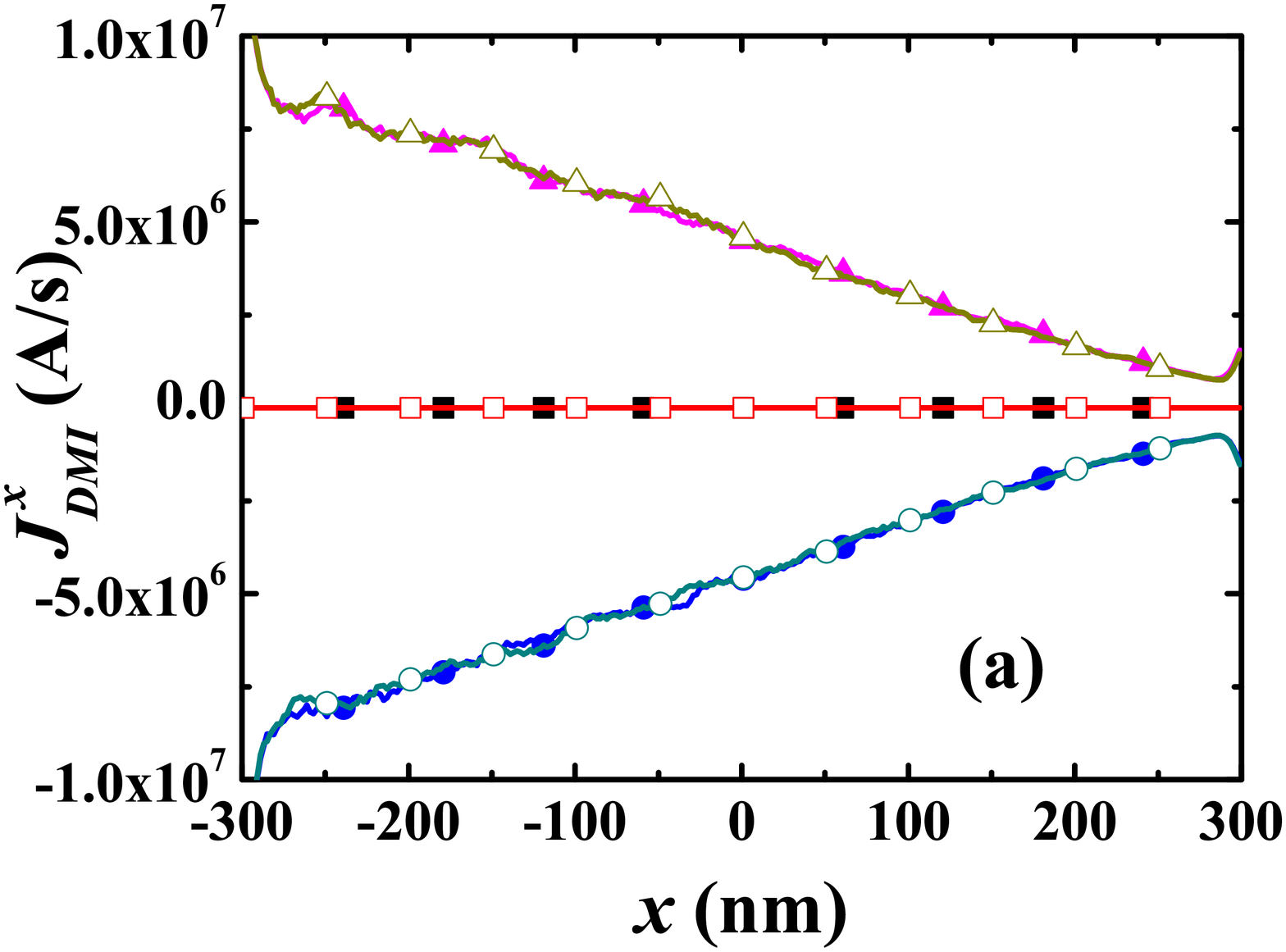}
\includegraphics[width=0.49\textwidth]{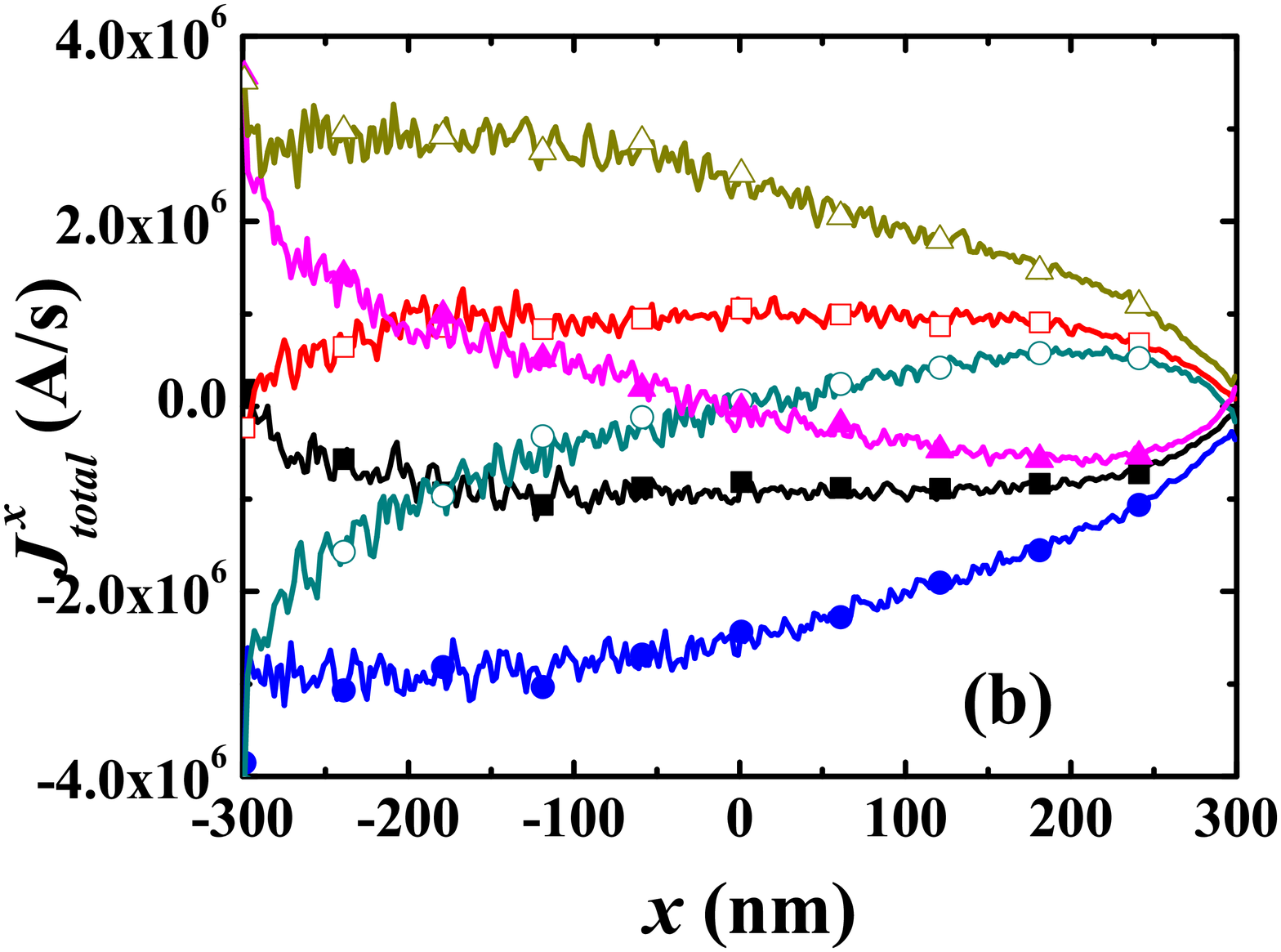}
\caption{The DMI magnonic spin current $J^x_{DMI}$ (a) and the total magnonic spin current $J_{\rm total}$ along the x axis in the nanowire. Values of the parameters $D = 0$ (squares), $D = +1.58$ mJ/m$^2$ (circles) and $D = -1.58$ mJ/m$^2$ (triangles) in the uniform magnet with positive (solid dots) or negative (open dots) magnetization. Implemented thermal bias $dT=0.03$K/nm. }
\label{fig_2}
\end{figure}

The direction of the magnetization current induced by the DMI,  $J^x_{DMI} = -\frac{1}{2}\rho l_{DMI}$ (cf.  (Eq. (\ref{eq_4}) depends on the sign of the DMI constant $D$ while it is independent of the magnetization orientation,
see Fig. \ref{fig_2}(a).
For example, if $D > 0$,
$J^x_{DMI}$ is positive and flows along the $x$ axis, while it is reversed  if $D$ is negative.
We also found that  the existence of DMI changes  the features of the exchange-induced current $J_A^x$, as shown in Fig. \ref{fig_1}(b), which is also reflected in the total magnetization current shown in Fig.5b.
When $D = 1.58 \cdot 10^{-3}$ J/m$^2$ and the magnetization is positive, $J_A^x$ is positive as well, which means that the exchange magnons flow towards the hot edge. As for the negative magnetization, $J_A^x$ is still positive and the exchange magnons flow towards the cold edge. Hence, the current density becomes larger compared to that for positive magnetization. For both positive and negative magnetization, the exchange spin current in major part of the wire is larger than for $D = 0$. When $D = -1.58 \cdot 10^{-3}$ J/m$^2$, the influence of DMI on $J_A^x$ is reversed.The observed phenomenon might be interesting for applications, e.g. for thermal diodes \cite{khomeriki}. By switching of the magnetization direction one can rectify total magnonic spin current.

In order to explain the influence of DMI on the exchange magnonic spin current, we analyze the dispersion relation and the spectral characteristics of the magnons contributing to the spin current in  both ($D\neq 0$) and  ($D= 0$) cases. The magnon dispersion relation and the attenuation length $\Lambda$ can be estimated as follows \cite{MoSe13}
\begin{equation}
\displaystyle \omega = \frac{2\gamma}{\mu_0 M_S}\left(\sqrt{\left(K_x+Ak^2+K_z\right)\left(K_x+Ak^2\right)}\pm Dk \right),
\label{eq_8}
\end{equation}
\begin{equation}
\displaystyle \Lambda = \frac{2\gamma}{\alpha \omega \mu_0 M_S}\frac{2kA \pm D(\frac{\omega}{M_S} \mp kD)}{K_x+K_{z}/2+k^2A}.
\label{eq_9}
\end{equation}
Here, $\omega$ is the magnon frequency, $k$ is the magnon wave-vector, and $(\pm)$ corresponds to the magnons propagating parallel or opposite to the direction of the local equilibrium magnetization. Thus,  Eq. (\ref{eq_8}) illustrates an asymmetry that occurs in the spin-wave dispersion relation due to the DMI.
Using Eqs. (\ref{eq_8}) and (\ref{eq_9}) the magnon attenuation length can be estimated directly. Results of our calculations (not shown) confirmed that the existence of the DMI slightly increases the magnon attenuation length. Due to the decrease of the magnon damping, the local magnon density is slightly increased compared to the case $D = 0$ (the increase of the local magnon density is shown in Fig. \ref{fig_1}(a)). The impact of the DMI and the dependence of the exchange magnon current on the magnetization direction, also can be interpreted in terms of the asymmetry of spin-wave dispersion relation. The velocity of the exchange magnons is related to the exchange interaction $v=2 l_A k$. Due to the asymmetry of the spin-wave dispersion relation, the wave-numbers  $k$ of magnons propagating in the opposite directions are different, $k_{(\pm)}$. Therefore, the  magnon velocities, $v_{(\pm)}=2 l_A k_{(\pm)}$, are different, too.
A sufficiently strong DMI can change the direction of the propagation of spin waves.

\begin{figure}[htb]
\centering
\includegraphics[width=0.49\textwidth]{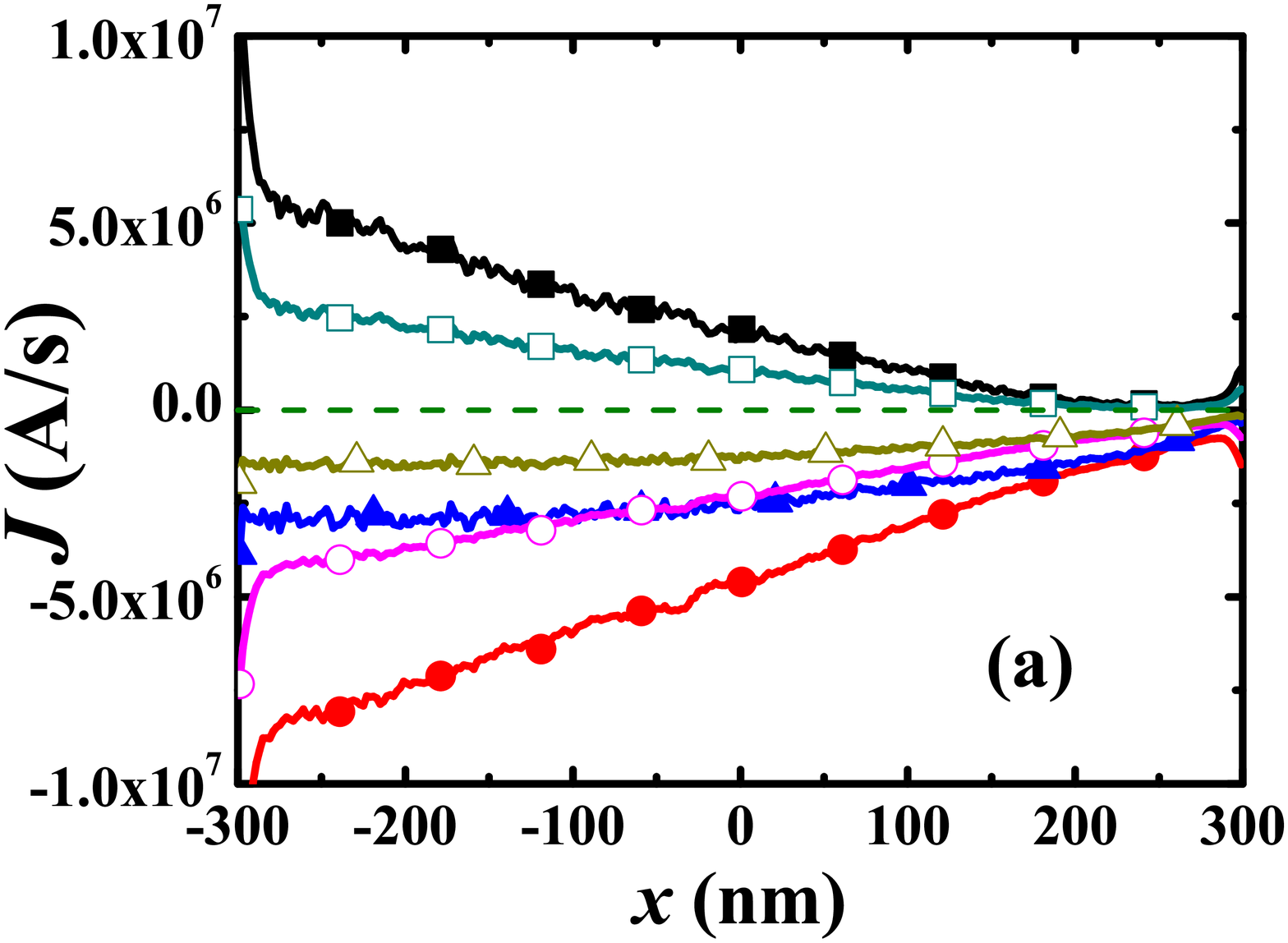}
\includegraphics[width=0.49\textwidth]{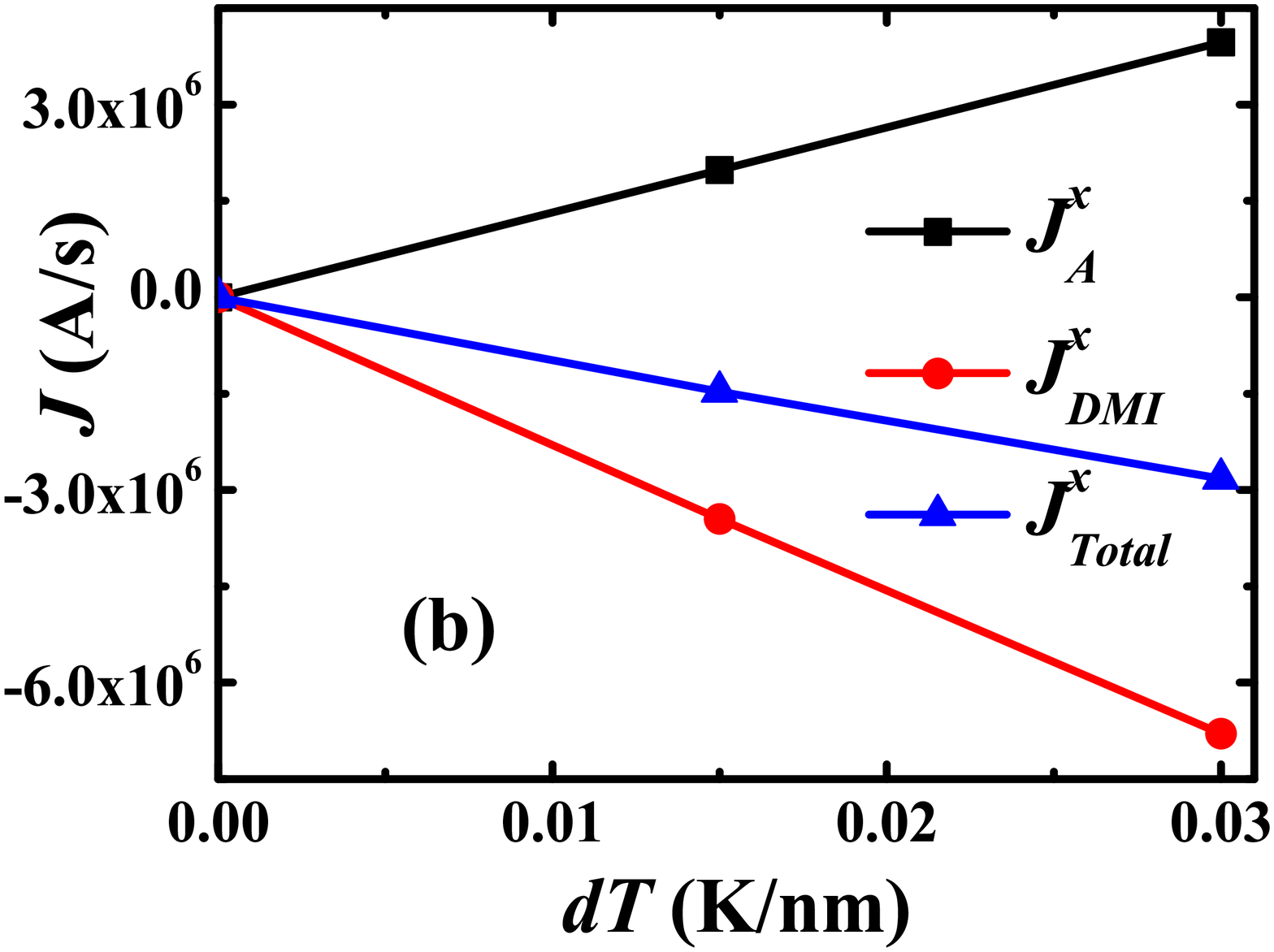}
\caption{(a) "The profiles of $J_{A}^{x}$  (squares), $J_{DMI}^{x}$  (circles) and  $J_{Total}^{x}$ (triangles) are plotted under different $\Delta T$ (0.03 K/nm (solid line with solid dots), 0.015 K/nm (solid line with open dots) and 0 K/nm (dash line without dots)) for the uniform positive magnetization and D = 1.58   10$^{-3}$ mJ/m$^{2}$".
(b)The averaged $J_{A}^{x}$  (squares), $J_{DMI}^{x}$  (circles) and  $J_{Total}^{x}$ (triangles) in the region of  300 nm $< x < 0$ nm as a function of the temperature gradient $\Delta T$ for the uniform positive magnetization and D = $1.58\cdot10^{-3}$ mJ/m$^{2}$.}
\label{fig 6}
\end{figure}

Another interesting effect of the DMI concerns the magnonic spin current at the edges.
Of key interest is the influence of the DMI on the magnon dispersion relation, Eqs. (\ref{eq_8}) and (\ref{eq_9}). At the edges of the nanowire,
 the magnons are reflected by the boundaries. After reflection the magnons change their velocity, $\vec{v}=-\vec{v}$, but not their spin (the scattering boundaries  do not discriminate spins in our case). Therefore,
  magnonic spin currents formed by the propagating and the reflected magnons  have opposite  signs and compensate each other in the vicinity of the boundaries.
  However, this holds only in absence of DMI. In fact, DMI changes the dispersion relation for magnons, Eqs. (\ref{eq_8}) and (\ref{eq_9}). The group velocities of the propagating and the reflected magnons turn different. Therefore, the two different spin currents formed by the propagating and the reflected magnons at the boundaries don't compensate each other. As  can be seen, this effect is stronger at the left edge of the nanowire Fig.\ref{fig_1} b, where the
temperature is higher, and therefore the concentration of magnons is also higher. Another interesting question concerns the dependence of magnonic spin current on the applied temperature gradient $\Delta T$.  Fig.\ref{fig 6} show the
$J_{A}^{x}$ and $J_{DMI}^{x}$ magnonic spin currents as a function of the temperature gradient.
 Our results confirm that both magnonic currents increase linearly with the applied temperature gradient and vanish when this temperature gradient is equal to zero.

\section{Exchange and DMI thermomagnonic torques}

Now, we consider thermomagnonic torques exerted on DWs. The magnonic spin current considered above generates torques. Since we decomposed the total spin current into exchange and DMI components,  we will discuss now the  corresponding torques.
To do this we write the total magnetization  $\vec{M}$ as a sum of the transversal and longitudinal components,
\begin{equation}
\label{longitudinal01}
 \displaystyle \vec{M}=M_{S0}\vec{m}_0+\vec{M}_{sw},
\end{equation}
where $M_{S0}=\sqrt{M^2_S-|M_{sw}|^2}$, and the local magnetization unit vector $\vec{m}_0$ describes the smoothly varying magnetic texture of the DW. After considering Eq. (\ref{longitudinal01}) and averaging over the fast oscillations in time, the expression of the exchange thermomagnonic torque,  $-l_A\vec{M}\times (\nabla^2 \vec{M})$, splits in two different parts: the magnonic adiabatic spin transfer torque, $-u\partial_x \vec{m}_0$,\cite{HiNo11,EtCh14,KiTs15}, and the exchange entropic torque, $l_A/2 \vec{m}_0\times (\partial_x \rho)(\partial_x\vec{m}_0)$.\cite{ScRi14,KiTs15} Here,  $u=l_A\left[\vec{m}_0\cdot \big \langle \vec{M}_{sw}\times \partial_x \vec{M}_{sw} \big \rangle\right]$ represents the exchange magnonic current, while $\rho=\langle |M_{sw}|\rangle^2$ stands for the average squared amplitude related to the magnon-number density. We adopt parametrization in terms of  spherical coordinates:
\begin{subequations}
\begin{align}
\displaystyle & \vec{m}_0=(\cos\theta,\sin \theta \cos \phi,\sin\theta \sin \phi), \\
 \displaystyle & \vec{M}_{sw}=M_u\vec{e}_{\theta}+M_v\vec{e}_{\phi}.
  \label{parametrization}
  \end{align}
\end{subequations}
After a little algebra we obtain
\begin{equation}
\label{torque01}
 \displaystyle u=-l_A \langle M_v\partial_x M_u - M_u\partial_x M_v \rangle,
\end{equation}
and
\begin{equation}
\label{torque01}
 \displaystyle \rho = \big \langle |M_u|^2+|M_v|^2 \big \rangle.
\end{equation}
Here $u$ recovers the expression for the exchange magnonic spin current $J_A^x$ obtained in the Cartesian coordinates Eq. (\ref{eq_2}).

A similar ansatz can be used for the DMI thermomagnonic torque. Our main interest is in the bulk DMI thermomagnonic torque $l_D\vec{M} \times \left(\nabla \times \vec{M} \right)$. We implement a parametrization in terms of  spherical coordinates Eq. (\ref{parametrization}) and  neglect the radial component of the DMI thermomagnonic torque coupled to the slow longitudinal magnetization component $\vec{m}_0$ (see Eq. (\ref{longitudinal01})). The time dependence of the angles in Eq. (\ref{parametrization}) is slower compared to the fast transversal magnetization components $M_u$ and $M_v$. Therefore on the characteristic time scale of $M_u$ and $M_v$ we consider angles as constant parameters. Due to  randomness, the average values of the transversal terms $M_u$ and $M_v$ are zero and only terms quadratic in $M_u$ and $M_v$  contribute. Therefore, after straightforward calculations  we obtain the following formula for the thermomagnonic torque:
\begin{equation}
\label{torque01}
 \displaystyle (l_D/4)\partial_x \rho \sin\theta \vec{e}_{\theta}+l_D M_u(\partial_xM_v)\sin \theta \vec{e}_{\phi}.
\end{equation}
The first term, $(l_D/4)\partial_x \rho \sin \theta \vec{e}_{\theta}$, called afterwards as the dissipative DMI torque, has certain similarity with the exchange entropic torque and depends on the magnon density gradient $\partial_x \rho$. The second term, $l_DM_u(\partial_x M_v)\sin \theta \vec{e}_{\phi}$, is similar to the DMI field-like or the reactive torque \cite{MaNd14,KoGu15}. Below it will be shown that this second term, $l_D M_u (\partial_x M_v)\sin \theta \vec{e}_{\phi}$,  is also similar to the magnonic momentum transfer DMI torque studied in Ref. \cite{WaAl15}. In the case of a circular symmetry, $\langle M_v \partial_x M_u \rangle = - \langle M_u \partial_x M_v \rangle$, this term  simplifies further to $Du/(2A)\sin \theta \vec{e}_{\phi}$. Both  DMI thermomagnonic torques are independent of the spatial variation of the local magnetization $\vec{m}_0$ and can exist in a homogeneous magnet, as well.

\begin{figure}[htb]
\centering
\includegraphics[width=0.49\textwidth]{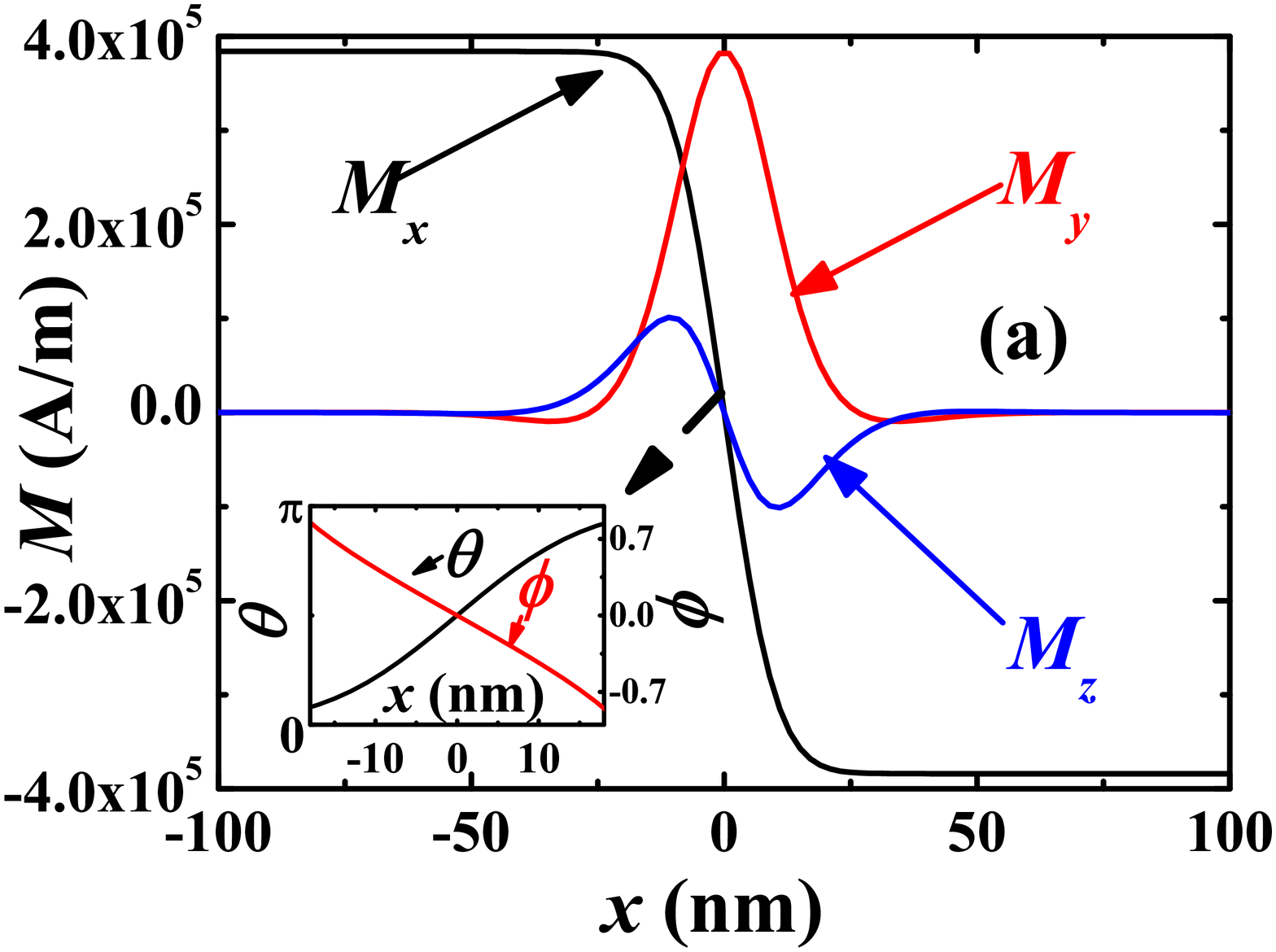}
\includegraphics[width=0.49\textwidth]{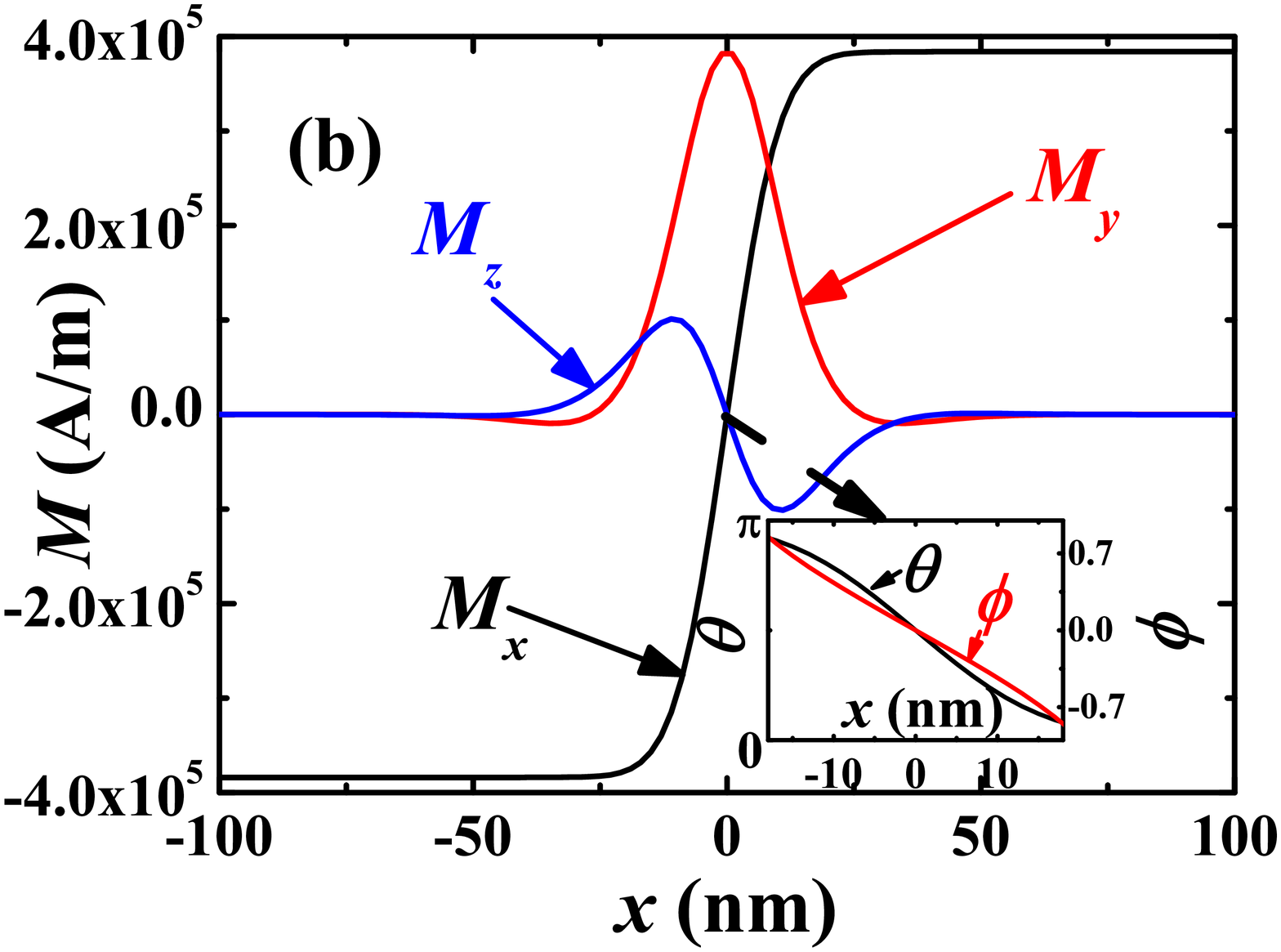}
\caption{Magnetization components of the static HH (a) and TT (b) DW profiles. The spatial distributions of spherical coordinates $\theta$ and $\phi$ in the DW region is shown in the insets. $D=$ $-1.58$ mJ/m$^2$.}
\label{fig_3}
\end{figure}

\section{Domain wall dynamics}

For a comprehensible study of the thermally activated DW motion, all possible effects related to the applied thermal gradient: such as exchange and DMI thermomagnonic torques, should be included. Recent calculations \cite{Arthur} reveal that under certain conditions two  scenarios for the DW motion are possible: If the DW width   is small, the DW is transparent for thermal magnons, i.e.   the magnons pass through the DW without a sizeable change in the magnon's momentum. Naturally, the magnonic spin current does not exert a magnonic pressure on the DW's surface, while the angular momentum is still transferred. In this case DW moves to the hot edge and this process is predetermined by the free energy of the DW \cite{ScRi14}. When the width of the DWs exceeds a critical value, the magnonic spin current is totaly reflected by DW. This leads to a strong recoil effect and a magnonic pressure. The DW moves to the cold edge. Here we study the dynamics of  thin DWs which are transparent for magnonic currents in the presence of DMI \cite{WaAl15}.

For the study of thermally activated motion of  DWs,  we consider the LLG equation supplemented by thermomagnonic torques
\begin{equation}
\displaystyle \partial_t \vec{M}_0 = - \gamma \vec{M}_0 \times \vec{H}_{{\rm eff}\, 0} + \alpha \vec{m}_0 \times \partial_t \vec{M}_0 + \vec{\tau}_{\rm exch} + \vec{\tau}_{\rm DMI}.
\label{eq_10}
\end{equation}
Here, $\vec{\tau}_{\rm exch}$ and $\vec{\tau}_{\rm DMI}$ are the exchange and the DMI thermomagnonic torques, respectively, $\vec{H}_{{\rm eff}\, 0}$ is the effective filed consisting of the exchange field, the bulk type DMI effective field, the uniaxial anisotropy field along the x axis, and the effective easy xy-plane anisotropy filed. Taking into account
Eq.(\ref{longitudinal01}), Eq.(\ref{parametrization}), and Eq.(\ref{torque01}), one can rewrite  Eq. (\ref{eq_10}) in  spherical coordinates as
\begin{equation}
\begin{split}
\displaystyle & M_{S0} (\partial_t \theta + \alpha \sin \theta \partial_t \phi) \\
\displaystyle &= \gamma H_{{\rm eff}\, 0 \phi}M_{S0} - u\partial_x \theta - \frac{l_A}{2} \partial_x \rho \sin \theta \partial_x \phi  + \frac{l_D}{4} \partial_x \rho \sin \theta,
\label{eq_11}
\end{split}
\end{equation}
\begin{equation}
\begin{split}
\displaystyle & M_{S0} (\sin \theta \partial_t \phi - \alpha \partial_t \theta) \\
\displaystyle &=- \gamma H_{{\rm eff}\, 0 \theta}M_{S0} - u \sin \theta \partial_x \phi + \frac{l_A}{2} \partial_x \rho \partial_x \theta  + \frac{Du}{2A} \sin \theta.
\label{eq_12}
\end{split}
\end{equation}
Here $H_{{\rm eff}\, 0 \theta}$ and $H_{{\rm eff}\, 0 \phi}$ are the components of the effective filed $\vec{H}_{{\rm eff}\, 0}$ in  spherical coordinates.

Hereafter the dynamics of the DW governed by the thermomagnonic torques is investigated based on the Eq. (\ref{eq_11}) and Eq. (\ref{eq_12}). We consider two different types of magnetic configurations for the DW: head-to-head (HH) and tail-to-tail (TT), see Fig. \ref{fig_3} (a) and (b) for details. Using two collective coordinates for the DW (first $q(t)$ characterizes the position and the second one is the tilt angle $\phi_1(t)$), we deduce the following profiles of the different DWs,
\begin{equation}
\displaystyle \theta=\arccos (p_1\tanh[(x-q(t))/\Delta]),
\label{profile01}
\end{equation}
and
\begin{equation}
\displaystyle \phi=\phi_1(t) + \phi_2(x-q(t)).
\label{profile02}
\end{equation}
Here, $p_1 = -1$ and $p_1 = 1$ correspond to the HH and TT walls, respectively, and $\Delta$ is the width of the DW. The DMI leads to a finite distortion of the DW's configuration. The distortion function $\phi_2$ is a linear function of $x-q$,  $\phi_2=(x-q)/\xi$ \cite{WaAl15}. Taking into account the static configuration of the DW we immediately find that $D/\xi >0$ is a positive constant.

Assuming the wall width $\Delta$ and the distortion $\phi_2$ to remain constant during the motion, we insert the DW profile into Eq. (\ref{eq_11}) and Eq. (\ref{eq_12}). After integration over the $x$ coordinate we obtain
\begin{equation}
\begin{split}
\displaystyle & \frac{p_1}{\Delta}\partial_t q + \alpha \partial_t \phi_1 - \frac{\alpha}{\xi}\partial_t q \\
\displaystyle &= -\frac{l_{\phi 2}\gamma K_z M_{S0}}{l_{\phi 1}\mu_0 M^2_S}\sin(2\phi_1)+\frac{p_1u}{\Delta M_{S0}} - \frac{l_A\partial_x \rho}{2\xi M_{S0}} + \frac{l_D \partial_x \rho}{4 M_{S0}},
\label{eq_13}
\end{split}
\end{equation}
\begin{equation}
\begin{split}
\displaystyle & -\frac{\alpha p_1}{\Delta}\partial_t q + \partial_t \phi_1 - \frac{1}{\xi}\partial_t q \\
\displaystyle & = \frac{l_{\phi 3}\gamma K_z M_{S0}}{l_{\phi 1}\mu_0 M^2_S}\sin(2\phi_1) - \frac{u}{\xi M_{S0}} - \frac{p_1 l_A\partial_x \rho}{2 \Delta M_{S0}} + \frac{D u}{2A M_{S0}}.
\label{eq_14}
\end{split}
\end{equation}
Here, the parameters
\begin{equation}
\begin{split}
\displaystyle & l_{\phi 1} = \int_{q-\phi \Delta/2}^{q+\pi \Delta /2}| \sin \theta | dx = 2.32 \Delta, \\
\displaystyle & l_{\phi 2} = \int_{q-\pi \Delta/2}^{q+\pi \Delta/2} \sin\theta \cos(2\phi_2)dx,
\label{Integral01}
\end{split}
\end{equation}
and
\begin{equation}
\displaystyle l_{\phi 3} = \int_{q-\pi \Delta/2}^{q+\pi \Delta/2} \cos\theta \sin \theta\sin(2\phi_2)dx,
\label{Integral02}
\end{equation}
depend on the DW structure and the DW distortion $\phi_2$. After a little algebra Eqs. (\ref{eq_13}) and (\ref{eq_14}) can be transformed to the form
\begin{equation}
\begin{split}
\displaystyle & \left(\frac{p_1l_{\phi 3}}{\Delta} - \frac{l_{\phi 2}}{\xi}\right)\partial_t q + \left(\alpha l_{\phi 3}+l_{\phi 2}\right)\partial_t \phi_1 \\
\displaystyle & - \left(\frac{p_1l_{\phi 2}}{\Delta} + \frac{l_{\phi 3}}{\xi}\right) \alpha \partial_t q = - \frac{u}{M_{S0}} \left(-\frac{p_1 l_{\phi3}}{\Delta} + \frac{l_{\phi 2}}{\xi}\right) \\
\displaystyle &  - \left(\frac{p_1l_{\phi 2}}{\Delta} + \frac{l_{\phi 3}}{\xi}\right) \frac{l_A \partial_x \rho}{2M_{S0}} + \frac{l_{\phi 2}Du}{2M_{S0}A} + \frac{l_{\phi 3}l_D \partial_x \rho}{4M_{S0}}.
\label{eq_15}
\end{split}
\end{equation}
In the absence of the DMI and DW distortion, i.e. $D = 0$, $l_{\phi 2} / \xi$ = 0 and $l_{\phi 3} = 0$, the motion of the DW is stipulated by the exchange entropic toque only, and the DW velocity is equal to $l_A \partial_x\rho/ (\alpha2M_{S0 })$ and $\partial_t\phi_1 = 0$.

The DMI  influences the DW motion. The contribution of the  adiabatic torque $-u(\partial \vec{m}_0/\partial x)$ in the steady velocity $\partial_t \phi_1 = 0$ of the DW  can be obtained from Eq. (\ref{eq_15}) and reads

\begin{equation}
\begin{split}
\displaystyle & v_{ad} = -u(-p_1l_{\phi 3}/\Delta + l_{\phi 2}/\xi)/(M_{S0}[(p_1l_{\phi 3}/\Delta-l_{\phi 2/\xi})- \\
\displaystyle & \alpha (p_1l_{\phi 2}/\Delta +l_{\phi 3}/\xi)]).
\label{eq_16}
\end{split}
\end{equation}

As one can see, the velocity $v_{ad}$ is proportional to the constant $-p_1l_{\phi 3}/\Delta + l_{\phi 2}/\xi$. For a stable DW structure, the factor $-p1l_{\phi 3}/\Delta + l_{\phi 2}/\xi$ is  small (see bellow) and hence is negligible. Therefore, we can argue that the adiabatic magnonic torque is almost irrelevant for the DW motion. In support of this statement we performed micromagnetic calculations and studied the motion of the DW induced solely by the adiabatic torque $u(\partial \vec{m}_0)/\partial x$ (not shown). We observed that in this case the velocity of the DW is really negligibly small.

For the velocity of the DW's steady motion, $\partial_t \phi_1 = 0$, driven  by thermomagnonic torques, we find
\begin{equation}
\begin{split}
\displaystyle v_{DW} = \frac{l_A \partial_x \rho}{2M_{S0} \alpha} - & \frac{l_D\Delta^2\partial_x \rho}{4\xi \alpha M_{S0}(1+ \Delta^2/\xi^2)} \\
\displaystyle & - \frac{Du\Delta}{2p_1\alpha M_{S0} A(1+ \Delta^2/\xi^2)}.
\label{eq_16}
\end{split}
\end{equation}
\begin{figure}[htb]
\centering
\includegraphics[width=0.49\textwidth]{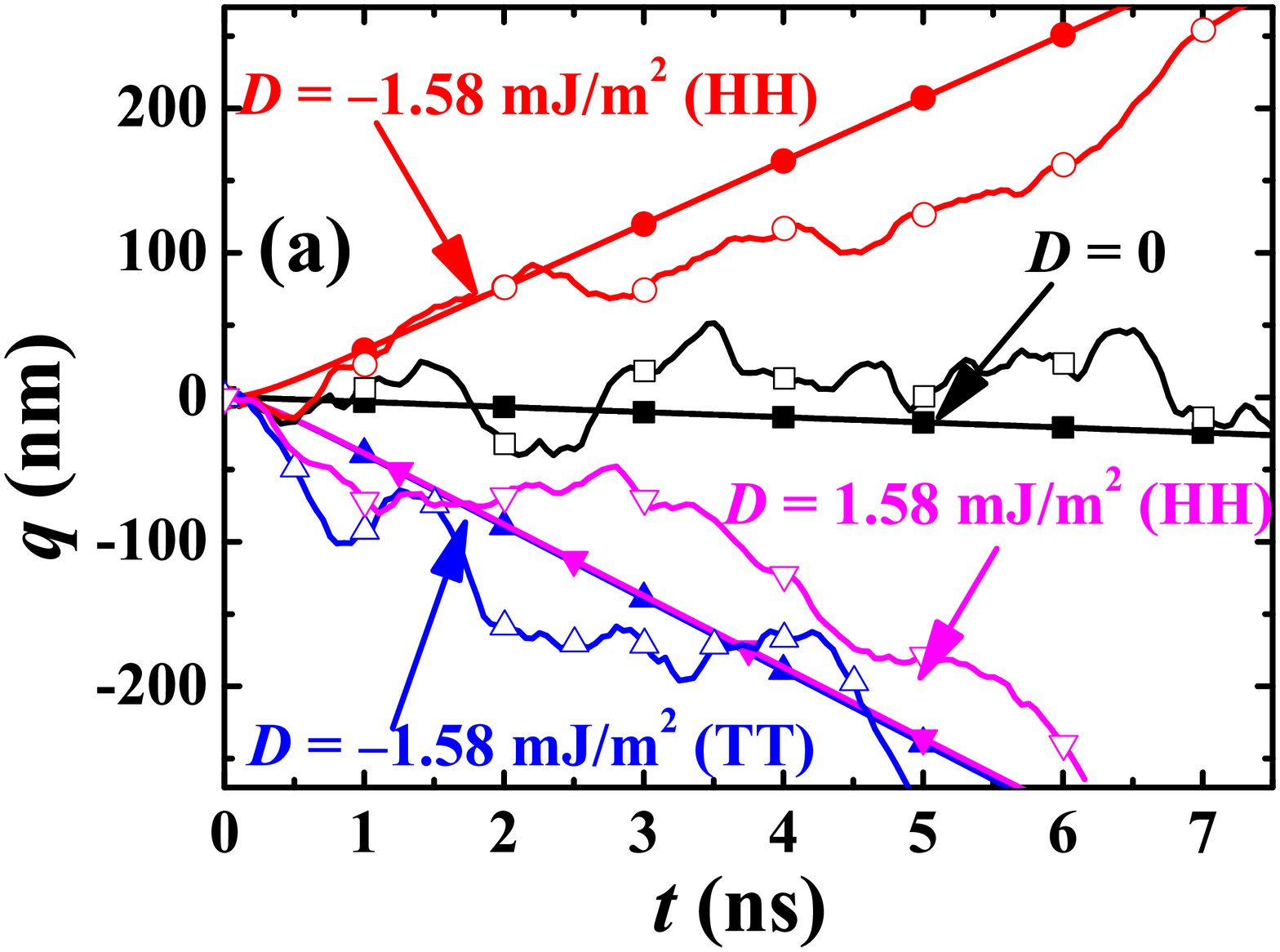}
\includegraphics[width=0.49\textwidth]{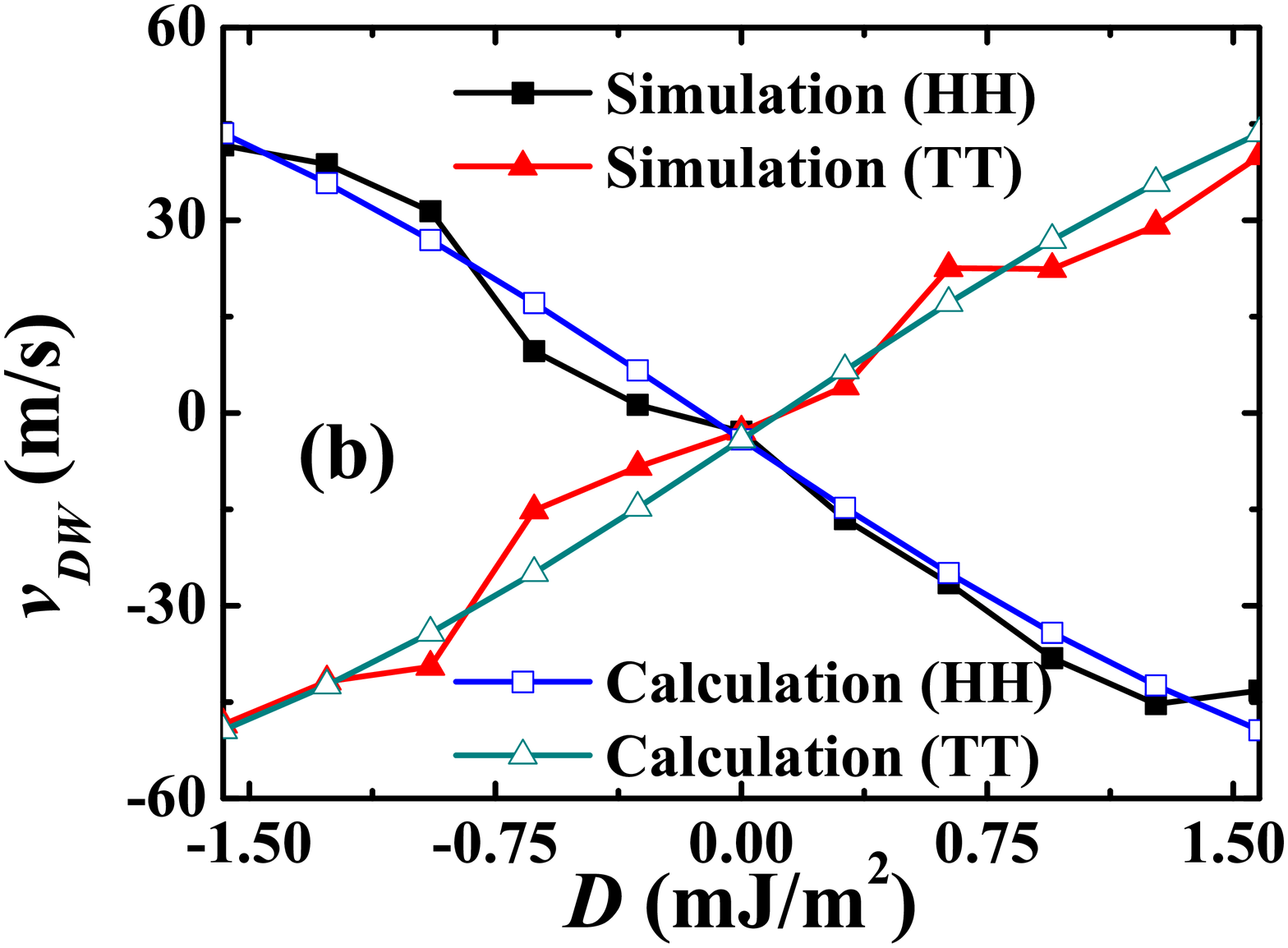}
\caption{(a) Numerically simulated (open dots) and analytically calculated (solid dots) domain wall displacement $q$ as a function of time $t$, induced by the temperature gradient $dT = 0.03$ K/nm  for $D = 0$ and HH wall (black squares), $D = -1.58$ mJ/m$^2$ and HH wall (red circles), $D = -1.58$ mJ/m$^2$ and TT wall (blue up-triangles) and $D = 1.58$ mJ/m$^2$ and HH wall (magenta down-triangles). (b) Analytically calculated (blue open square for HH wall and dark cyan open triangle for TT wall) and simulated (black solid square for HH wall and red solid triangle for TT wall) domain wall velocity $v_{dw}$ as a function of the DMI constant $D$.}
\label{fig_4}
\end{figure}
As we see from Eq. (\ref{eq_16}), the velocity of the DW is a sum of three distinct contributions: exchange entropic torque, the DMI entropic torque and the DMI field-like torque, while the magnonic adiabatic spin transfer torque is absent. The exchange entropic torque pushes the DW towards the hot edge. Thus, we recover the result of the previous studies \cite{ScRi14,WaWa14}. The effect of the DMI field-like torque on the DW's dynamics is equivalent to the effect of the effective magnetic field $H_x = D u / (2\gamma AM_{S0})$ applied along the x axis. Depending on the sign of $D$ and depending on the DW's structure, the DMI field-like torque can drag the DW in  both directions (opposite or parallel to the applied thermal bias). The DMI field-like torque is analogous to the DMI magnonic momentum transfer torque studied in Ref. \cite{WaAl15}. As evident, the DMI entropic torque is quite susceptible to the DMI constant $D$ and to the DW distortion parameter $\xi$. This forces DW to move to the cold edge if $ l_D/\xi> 0$.

The material parameters used in the numerical calculations read: $M_S = 3.84 \cdot 10^5$ A/m, $A = 8.78 \cdot 10^{-12}$ J/m, $K_x = 1 \cdot 10^5$ J/m$^3$, $K_z = 2 \cdot 10^5$ J/m$^3$ and $\alpha= 0.05$. The strength of the DMI  is varied within the interval  $-1.58$ mJ/m$^2$ $\leq D\leq$  $1.58$ mJ/m$^2$d, $u$ and $\rho$ are determined from the simulation results via Eq. (\ref{eq_5}). The DW structure parameters $\Delta$ and $1/\xi$ are determined from the formed stable DW structure see Fig. \ref{fig_3}. As we see the DMI induced DW distortion $\phi$ is  a linear function of $x$. Its slope $1/\xi$ is negative for the both the HH and the TT configuration of the DWs when $D=$ $-1.58$ mJ/m$^2$. Apart from this,  we observed (not shown) that the strength of the slope $|1/\xi|$ decreases with the decrease of $|D|$ and the sign of $1/\xi$ changes with the DMI constant $D$.

The wall displacement $q(t)$ estimated from Eqs. (\ref{eq_13}) and (\ref{eq_14}) is shown in Figure. \ref{fig_4}(a) for the applied thermal bias $dT=$ $0.03$ K/nm. In the absence of the DMI ($D = 0$), the exchange entropic torque dominates and the DW moves toward the hot region ($-x$ direction). The DMI field-like torque may be much larger than the exchange and the DMI entropic torques. Therefore, the DMI field-like torque can enhance substantially the  DW speed and even switch the direction of the DW motion. According to Eq. (\ref{eq_16}), for $D > 0$ and TT wall (or $D < 0$ and the HH wall) the effect of the DMI field-like torque is opposite to the exchange entropic torque. By changing the sign of $D$ or DW structure, the effect of the DMI field-like torque and the direction of the wall motion can be reversed. Partial contributions of the different torques to the total speed of the DW have been extracted from Eq. (\ref{eq_16}). Assuming $|D| = 1.58 \cdot 10^{-3}$ J/m$^2$, for the exchange entropic torque, the DMI entropic torque and the DMI field-like torque we deduce $|l_A\partial_x\rho /(2M_{S0}\alpha)|$= 4.1 m/s, $|l_D \Delta^2 \partial_x \rho /(4\xi \alpha M_{S0} (1+\Delta^2/\xi^2))|= 1.2$ m/s and $|Du\Delta /(2p_1\alpha M_{S0}A(1+\Delta^2/\xi^2))|= 46.5$ m/s. Apparently the effect of the DMI field-like torque is much larger than other torques. Besides, in Eq. (\ref{eq_15}), $|p_1l_{\phi 3}/\Delta - \l_{\phi 2}/\xi$= 0.007 and $|v_{ad}|$ = 0.1 m/s. Thus, the effect of the magnonic adiabatic torque is weak enough and can be neglected safely.
\begin{figure}[htb]
\centering
\includegraphics[width=0.49\textwidth]{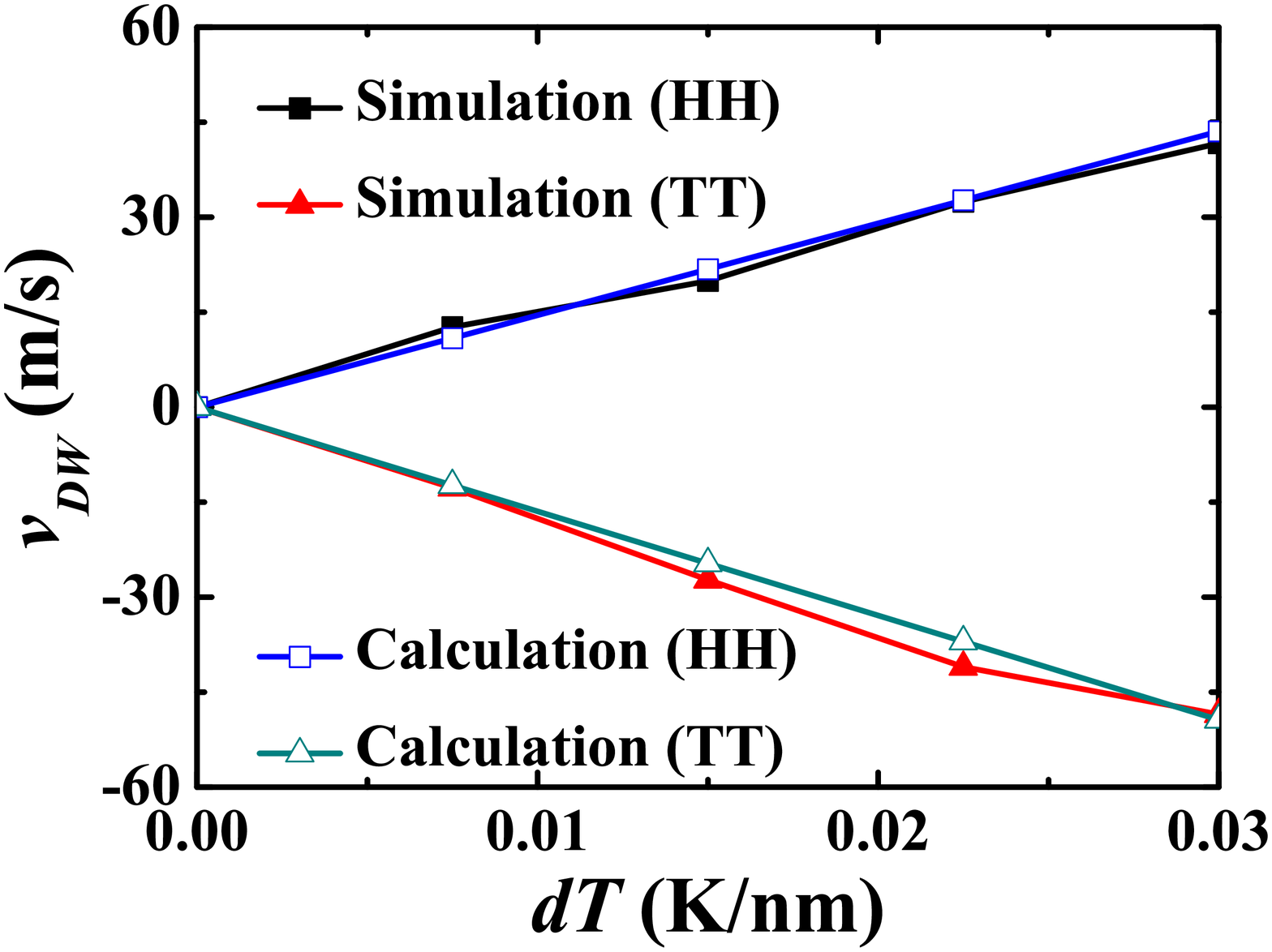}
\caption{ Analytically calculated (blue open square for HH wall and dark cyan open triangle for TT wall) and simulated (black solid square for HH wall and red solid triangle for TT wall) domain wall velocity $v_{dw}$ as a function of the temperature gradient $dT$ for $D = -1.58$ mJ/m$^2$.}
\label{fig_5}
\end{figure}

The connection between the DW velocity $v_{DW}$ and the DMI constant $D$ calculated from Eq. (\ref{eq_16}) for the HH and the TT DW configurations is plotted in Fig. \ref{fig_4}(b). With the decrease of $D$ from $1.58 \cdot 10^{-3}$ J/m$^2$ to $-1.58 \cdot 10^{-3}$ J/m$^2$, the HH DW velocity gradually switches from negative (\emph{-x})to the positive (\emph{x}) direction. In particular this happens when $D$ approaches  $D_{c}=-0.3 \cdot 10^{-3}$ J/m$^2$.

Apart from the analytical estimations, also micromagnetic numerical calculations have been performed. The results for the thermally assisted DW motion are shown in Fig. \ref{fig_4}(a) and \ref{fig_4}(b). Analytical estimations and exact micromagnetic calculations show a good agreement. Slight difference is caused by thermal fluctuations. We note that the DW structure is stable and the Walker breakdown is not observed. Moreover, the DW motion with DMI can be reasonably enhanced by increasing the thermal gradient $dT$. The linear dependence between the DW velocity $v_{DW}$ and the temperature slope $dT$ can be  observed clearly in the  results shown in Fig. \ref{fig_5}.

\section{Conclusions}
We have studied  the thermally activated DW motion in magnetic insulators. Our interest was especially focused on the effect of DMI and thermomagnonic torques on the DW motion. The thermally assisted DW motion is driven by the thermomagnonic spin current. Usually, the magnonic spin current is attributed to the exchange interaction only. Here, in addition to the exchange magnonic spin current,  the exchange adiabatic, and the entropic spin transfer torques we have also studied the DMI induced magnonic spin current, thermomagnonic DMI field-like torque, and the DMI entropic toque. Analytical estimations are supported by  numerical calculations. We have observed a dominant role of the DMI field-like torque (DMI momentum transfer torque). For a large DMI constant $D>0.3\cdot10^{-3}$ J/m$^{2}$, the influence of the DMI field-like torque is stronger compared to the DMI entropic torque and the exchange entropic torque.

Tuning   the DMI strength,  the DW speed  can be changed as well as  the direction of the DW motion. Analytical estimations are in a good agreement with the micromagnetics simulations. We have also observed that DMI not only contributes to the total magnonic spin current, but depending on the orientation of the steady state magnetization, the DMI surprisingly modifies the exchange magnonic spin current, a  phenomenon that might be exploited in  caloritronics. By switching the magnetization direction one
can rectify the total magnonic spin current.  The DMI  is found to  influence substantially  the geometry and  the shape of the DWs with the  DWs being  oriented parallel to the applied thermal bias. In this case instead of exerting pressure on DWs, the magnonic current
drags the DW. Furthermore,  we found that the magnonic current smoothes the magnetic texture.

\section{Acknowledgements}
This work is supported by the Deutsche Forschungsgemeinschaft under grants BE 2161/5-1 and SFB 762.
Guang-hua Guo acknowledge financial support from National Natural Science Foundation of China (No. 11374373), Doctoral Fund of Ministry of Education of China (No. 20120162110020) and the Natural Science Foundation of Hunan Province of China (No. 13JJ2004). VD and JB acknowledge support from  the National Science Center in Poland through the Project
No. DEC-2012/04/A/ST3/00372.



\end{document}